\documentclass[3p,times,twocolumn]{elsarticle}

\usepackage{ecrc}

\volume{00}

\firstpage{1}

\journalname{Nuclear and Particle Physics Proceedings}

\runauth{}

\jid{nppp}

\jnltitlelogo{Nuclear and Particle Physics Proceedings}

\usepackage{amssymb}

 \usepackage{lineno}

\usepackage[figuresright]{rotating}

\begin{document}

\begin{frontmatter}



\dochead{}

\title{Measurements of the suppression and anisotropy of heavy-flavour particles in Pb--Pb collisions at $\sqrt{s_\mathrm{NN}}$= 2.76 TeV with ALICE}


\author{Andrea Dubla, for the ALICE collaboration}

\address{GSI Helmholtz Centre for Heavy Ion Research, Planckstra{\ss}e 1 - 64291 Darmstadt, Germany}

\begin{abstract}

Heavy quarks, i.e. charm and beauty, are produced on a shorter time scale with respect to the strongly-interacting matter produced in high-energy heavy-ion collisions. Therefore, they are unique probes to study the mechanisms of parton energy loss, hadronisation and thermalization in the hot and dense state of matter.
The nuclear modification factor ($R_{\rm AA}$) and the elliptic flow ($v_{2}$) are two of the main experimental observables that allow us to investigate the interaction strength of heavy quarks with the medium.
The most recent results on heavy-flavour production and elliptic flow measured by the ALICE collaboration in Pb--Pb collisions at $\sqrt{s_\mathrm{NN}}$  = 2.76 TeV will be discussed.

\end{abstract}

\begin{keyword}
ALICE \sep heavy-flavour \sep heavy-ion

\end{keyword}

\end{frontmatter}

\section{Introduction}
The main goal of the ALICE \cite{ALICE2} experiment is to study strongly-interacting matter at high energy density and temperature reached in ultra-relativistic heavy-ion collisions at the Large Hadron Collider (LHC).
In such collisions a deconfined state of quarks and gluons, the Quark-Gluon Plasma (QGP), is expected to be formed.
Due to their large masses, heavy quarks, i.e. charm ($c$) and beauty ($b$) quarks, are produced at the early stage of the collision, almost exclusively in hard partonic scattering processes. Therefore, they interact with the medium in all phases of the system evolution, propagating through the hot and dense medium and loosing energy via radiative~\cite{Radiativea} and collisional~\cite{Colla} scattering processes. Heavy-flavour hadrons and their decay products are thus effective probes to study the properties of the medium created in heavy-ion collisions. 
The nuclear modification factor $R_\mathrm{AA}(p_\mathrm{T}) = \frac{1}{<T_\mathrm{AA}>} \frac{dN_\mathrm{AA}/d\it{p}_{T}}{d\sigma_\mathrm{pp}/{dp_\mathrm{T}}}$ of heavy-flavour hadrons (and their decay leptons) is well established as a sensitive observable to study the interaction strength of hard partons with the medium. 
Further insight into the medium properties is provided by the measurement of the anisotropy in the azimuthal distribution of particle momenta, that is characterized by the second Fourier coefficient \mbox{$v_{2} = < \cos[2(\varphi - \psi_{2})] >$}. 
At low $p_\mathrm{T}$, the $v_{2}$ of heavy-flavour hadrons is sensitive to the degree of thermalization of charm and beauty quarks in the deconfined medium and to different hadronisation mechanisms, namely the fragmentation in the vacuum and the coalescence in the medium~\cite{Coale}. At higher $p_\mathrm{T}$, the measurement of $v_{2}$ carries information on the path-length dependence of in-medium parton energy loss~\cite{highptv2}.
The measurement of heavy-flavour $v_{2}$ offers a unique opportunity to test whether also quarks with large mass participate in the collective expansion dynamics and possibly thermalize in the QGP.

Charm and beauty production was measured with ALICE in Pb--Pb collisions at $\sqrt{s_\mathrm{NN}}$  = 2.76 TeV using electrons at mid-rapidity and muons at forward rapidity from semi-leptonic decays of heavy-flavour hadrons and fully reconstructed D-meson hadronic decays. 
D mesons were reconstructed at mid-rapidity ($|y|$ $<$ 0.5) via their hadronic decay channels: ${\rm D}^{0} \rightarrow K^{-}\pi^{+}$, ${\rm D}^{+}\rightarrow K^{-}\pi^{+}\pi^{+}$, ${\rm D}^{*+} \rightarrow {\rm D}^{0}\pi^{+}$ and ${\rm D}^{+}_{s} \rightarrow \phi \pi^{+} \rightarrow K^{-}K^{+}\pi^{+}$ and their charge conjugates. 
The electron identification in the mid-rapidity region \mbox{ ($|y|$ $<$ 0.8)} was based on the d{\it E}/d{\it x} in the TPC. In the low $p_\mathrm{T}$ intervals ($p_\mathrm{T}$ $<$ 3 GeV/{\it c}), where the $K^{\pm}$, proton and deuteron Bethe-Bloch curves cross that of the electron, the measured time-of-flight in TOF and the energy loss in the ITS were employed in addition. At higher $p_\mathrm{T}$, the ratio of the energy deposited in the ElectroMagnetic Calorimeter (EMCal) and the momentum measured with the TPC and ITS, which is close to unity for $\mathrm{e}^{\pm}$, was used to further reject hadrons. Muon tracks were reconstructed in the Forward Muon Spectrometer \mbox{(-4 $<$ $\eta$ $<$ -2.5)}.
\label{}


\section{Nuclear modification factor}  \label{}

\begin{figure}[htb]
\centering
\includegraphics[height=2.8in]{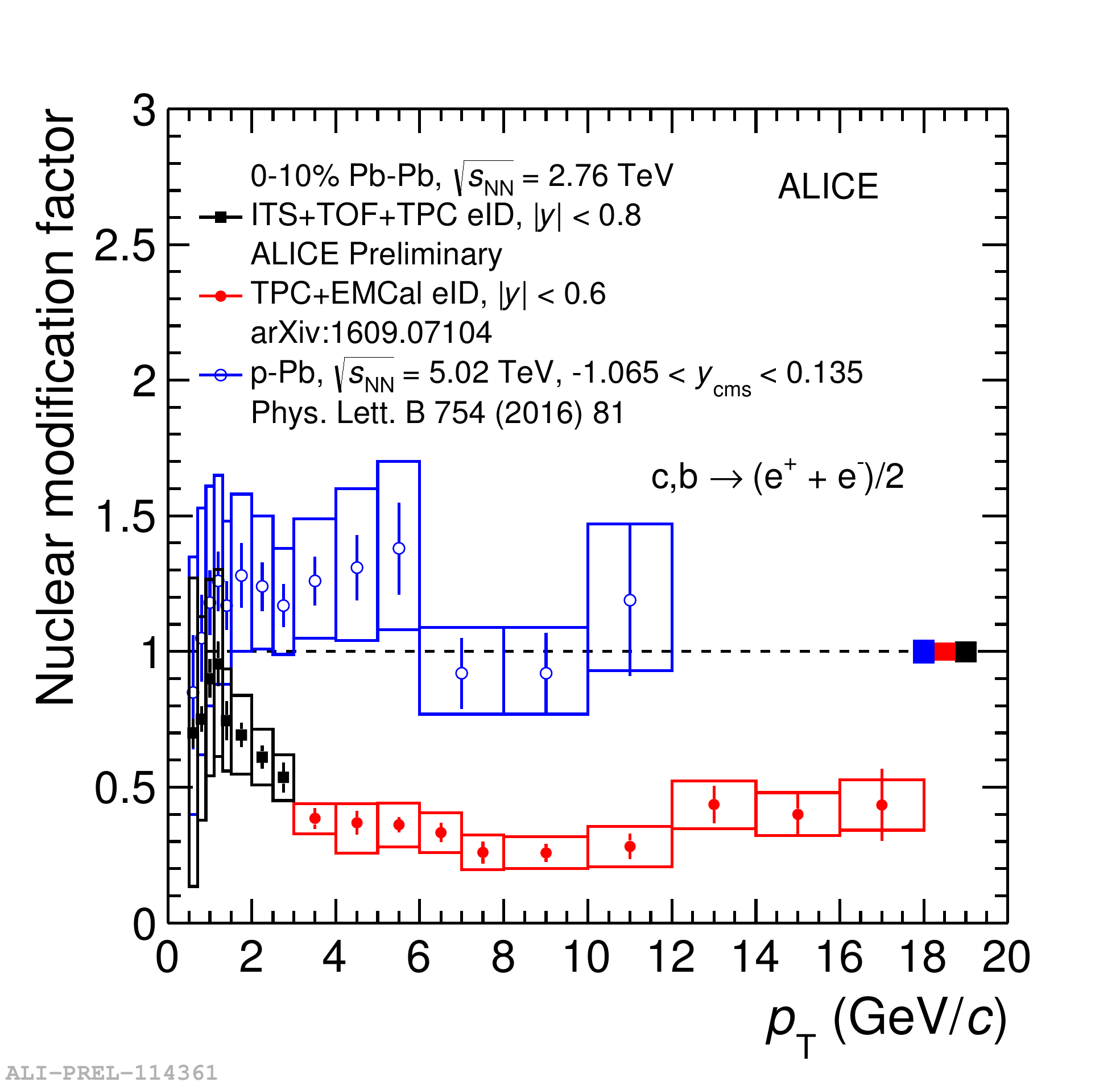}
\caption{Heavy-flavour decay electrons $R_\mathrm{AA}$ as a function of $p_\mathrm{T}$ in Pb--Pb collisions at $\sqrt{s_\mathrm{NN}}$ = 2.76 TeV in the 0-10\% centrality class and in p--Pb collisions at $\sqrt{s_\mathrm{NN}}$ = 5.02 TeV. }
\label{fig:figure1}
\end{figure}

The ALICE collaboration measured the $R_\mathrm{AA}$ of open heavy-flavour hadrons via their hadronic and semi-leptonic decays in Pb--Pb collisions at $\sqrt{s_\mathrm{NN}}$ = 2.76 TeV~\cite{MuonRAA, Shingopapaper, DMesonRPBPB}.
Figure~\ref{fig:figure1} shows the nuclear modification factor of electrons from heavy-flavour hadron decays as a function of $p_\mathrm{T}$ in central (0-10\%) Pb--Pb and in p--Pb collisions~\cite{Shingopapaper, HFERpPb}. A strong suppression is observed in central Pb--Pb collisions in the $p_\mathrm{T}$ range 3-18 GeV/$c$. The $R_\mathrm{AA}$ is observed to increase when moving to low $p_\mathrm{T}$ reaching the unity within the statistical and systematic uncertainty. The low-$p_\mathrm{T}$ measurement is very important to test the binary scaling of the heavy-flavour production and to measure the total $c\bar{c}$ cross section in Pb--Pb collisions. 
An $R_\mathrm{pPb}(p_\mathrm{T})$, also shown in Fig.~\ref{fig:figure1}, consistent with unity was measured~\cite{HFERpPb}, confirming that the suppression observed in central \mbox{Pb--Pb} collisions is predominantly induced by final-state effects due to charm quark energy loss in the medium. 

\begin{figure}[htb]
\centering
\includegraphics[height=2.6in]{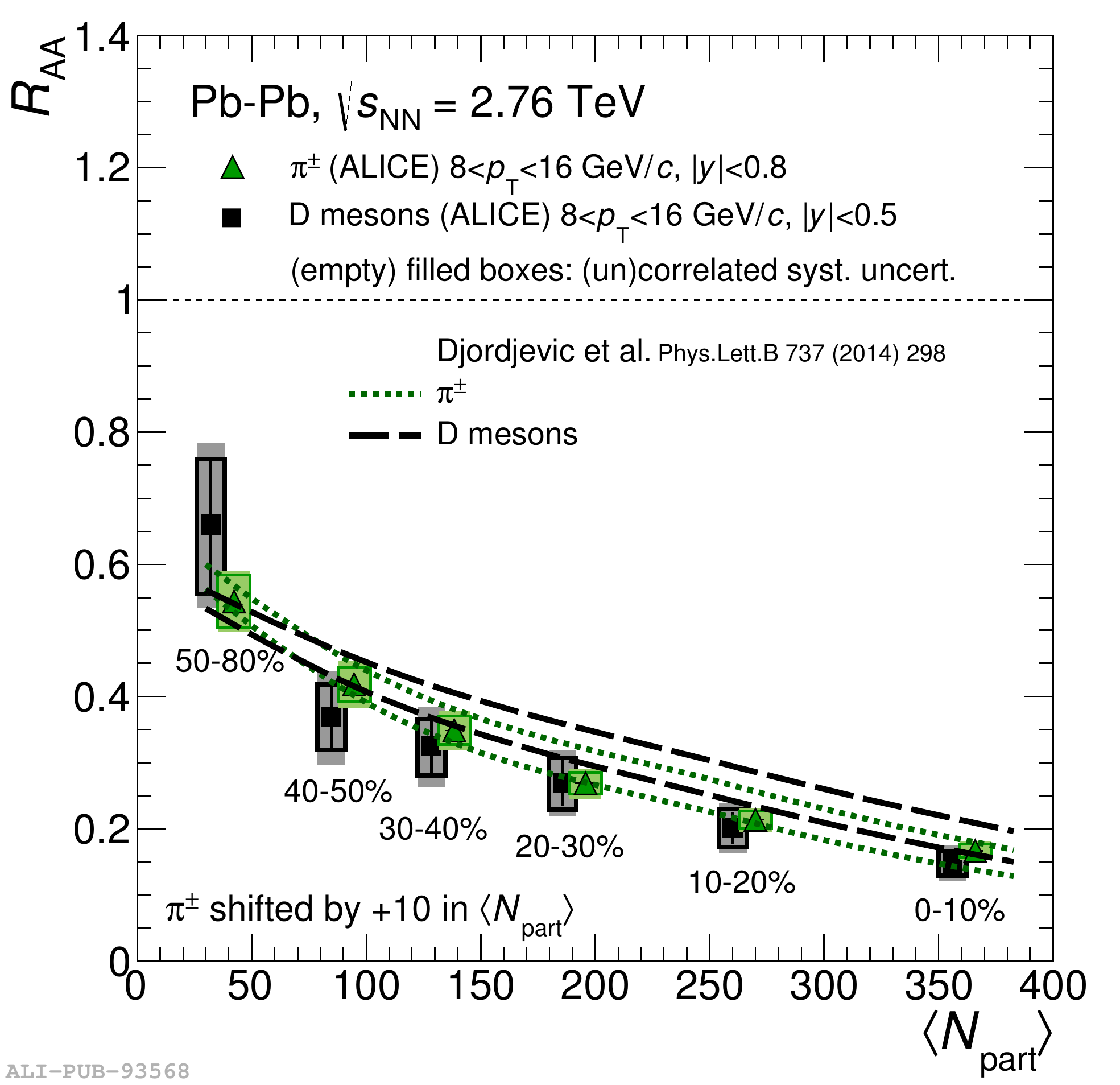}
\caption{Comparison of D mesons and charged pions $R_\mathrm{AA}$ measurements as a function of centrality~\cite{DmesonRAACent, RAAPion} and with the model calculations~\cite{Djordjevic} including radiative and collisional energy loss.}
\label{fig:figure2}
\end{figure}

Due to the QCD nature of parton energy loss, quarks are predicted to lose less energy than gluons (that have a larger colour coupling factor). In addition, the dead-cone effect \cite{DeadCone} is expected to reduce the energy loss of massive quarks with respect to light quarks.
Therefore, if we consider only energy-loss effect, a hierarchy in the $R_\mathrm{AA}$ is expected to be observed when comparing the mostly gluon-originated light-flavour hadrons (e.g. pions) to D and to B mesons \cite{DmesonRAACent}: $R_{\rm AA}({ \pi})$ $<$ $R_{\rm AA}({\rm D})$ $<$ $R_{\rm AA}({\rm B})$. The measurement and comparison of these different medium probes provides a unique test of the colour-charge and mass dependence of parton energy loss.
In Fig. \ref{fig:figure2} the nuclear modification factor of prompt D mesons in the transverse momentum region 8 $<$ $p_\mathrm{T}$ $<$ 16 GeV/{\it c} is shown as a function of centrality in comparison with the $R_{\rm AA}({ \pi})$. Less suppression is observed moving from central to semi-central collisions, since the medium formed in peripheral collisions should be less dense with respect to the one formed in a central collision. The D-meson $R_{\rm AA}({\rm D})$ is compatible with that of charged pions and charged particles within uncertainties \cite{DMesonRPBPB, DmesonRAACent, RAACH, RAAPion}. The consistency the two measurements, , which at first sight disagrees with the hierarchy of the $R_\mathrm{AA}$ mentioned above, is rather described by a model including mass-dependent energy loss, different shape of the parton $p_\mathrm{T}$ spectra and different parton fragmentation functions~\cite{Djordjevic}.

\begin{figure}[htb]
\centering
\includegraphics[height=2.6in]{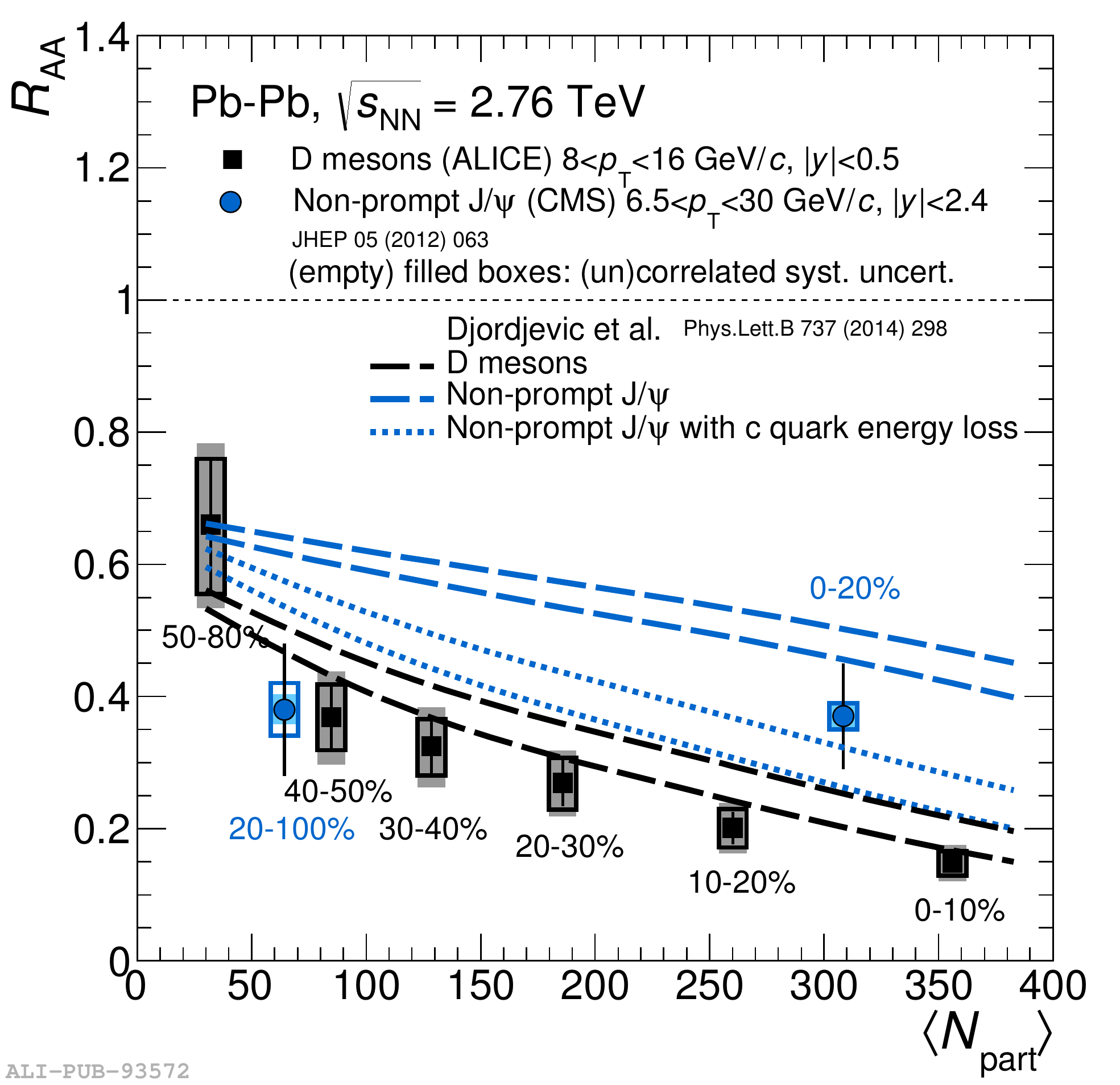}
\caption{Comparison of the $R_\mathrm{AA}$ measurements as a function of centrality for the D mesons in 8 $<$ $p_\mathrm{T}$ $<$ 16 GeV/{\it c}~\cite{DmesonRAACent} and non-prompt J/$\psi$ mesons in 6.5 $<$ $p_\mathrm{T}$ $<$ 30 GeV/{\it c}~\cite{RAACMS}.}
\label{fig:figure3}
\end{figure}

Finally the centrality dependent D-meson $R_\mathrm{AA}$ is compared with the $R_\mathrm{AA}$ of non-prompt J/$\psi$ measured by the CMS collaboration \cite{RAACMS} in Fig.~\ref{fig:figure3}, and an indication of a different suppression for the most central collisions is found as expected from the anticipated quark-mass dependence of the energy loss. \newline The two measurements are described by the predictions based on a pQCD model including mass-dependent radiative and collisional energy loss ~\cite{Djordjevic}.
In this model the difference in the $R_\mathrm{AA}$ of charm and beauty mesons is mainly due to the mass dependence of quark energy loss, as demonstrated by the difference in the  predictions when the non-prompt J/$\psi$ $R_\mathrm{AA}$ is calculated assuming that $b$ quarks suffer or not the same energy loss as $c$ quarks.


\begin{figure}[htb]
\centering
\includegraphics[height=2.6in]{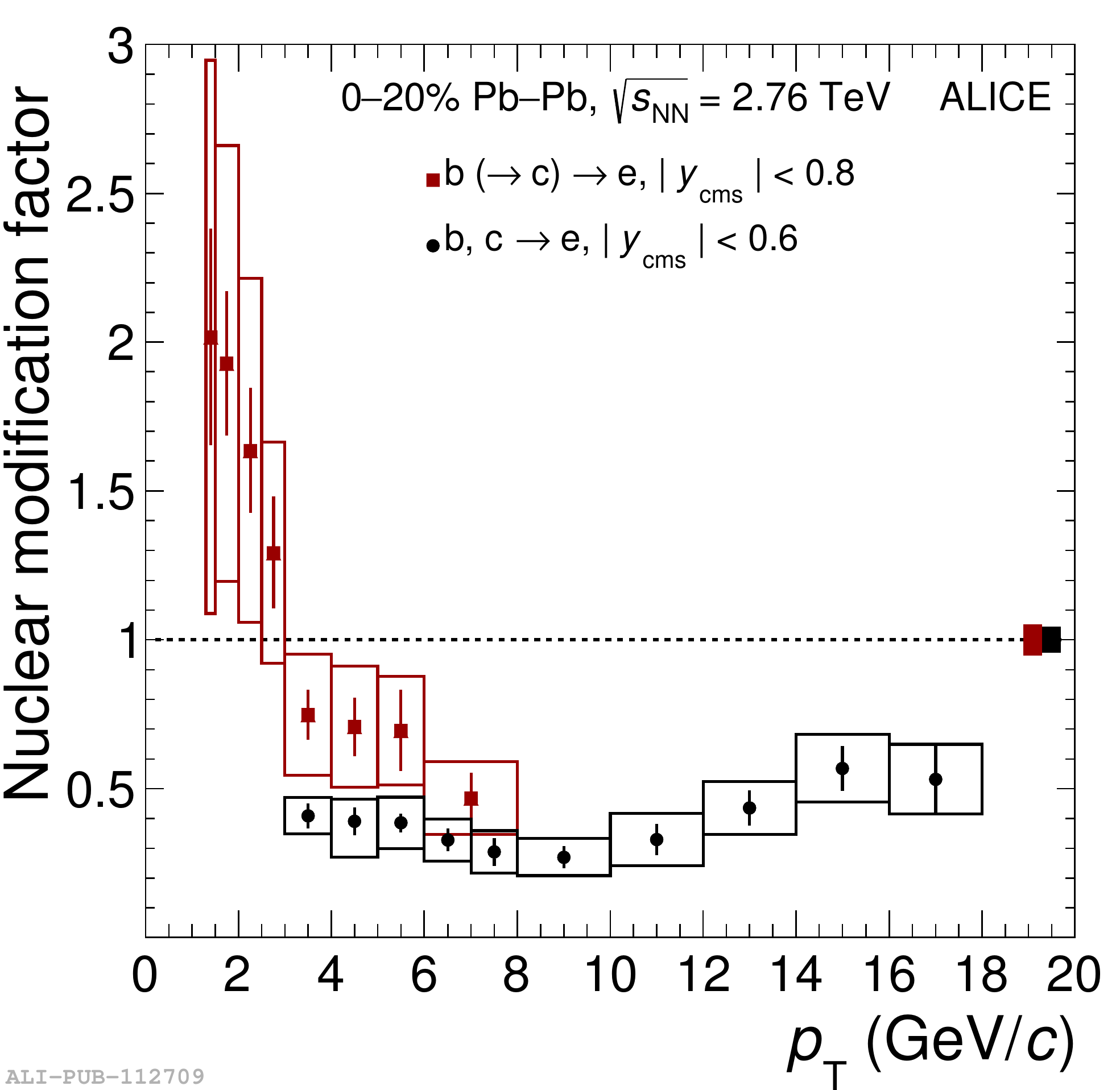}
\caption{$R_\mathrm{AA}$ of electrons from beauty-hadron decays together with the corresponding result for beauty- and charm-hadron decays for the 20\% most central Pb--Pb collisions.}
\label{fig:figure4}
\end{figure}

A comparison of the $R_\mathrm{AA}$ of electrons from beauty-hadron decays~\cite{Btoe} with the one from charm- plus beauty-hadron decays~\cite{Shingopapaper} is shown in Fig.~\ref{fig:figure4} for the 20\% most central Pb--Pb collisions. The contribution to the heavy-flavour decay electron yield due to beauty-hadron decays was extracted by means of a fit to the electron impact parameter distribution. The electron sources were included in the fit through templates obtained from simulations. 
The results agree within uncertainties at high $p_\mathrm{T}$, where the beauty contribution is larger than the charm contribution~\cite{Btoepp}. \newline
In the $p_\mathrm{T}$ interval 3--6 GeV/$c$, the $R_\mathrm{AA}$ for electrons from beauty-hadron decays is about 1.3$\sigma$ less than the one of semi-inclusive heavy-flavour hadron decay electrons . This difference is consistent with the ordering of charm and beauty suppression seen in the comparison between prompt D meson~\cite{DmesonRAACent} and non-prompt J/$\psi$ \cite{RAACMS}.

\section{Elliptic flow}  \label{}

\begin{figure*}[tp]
\includegraphics[height=2.in]{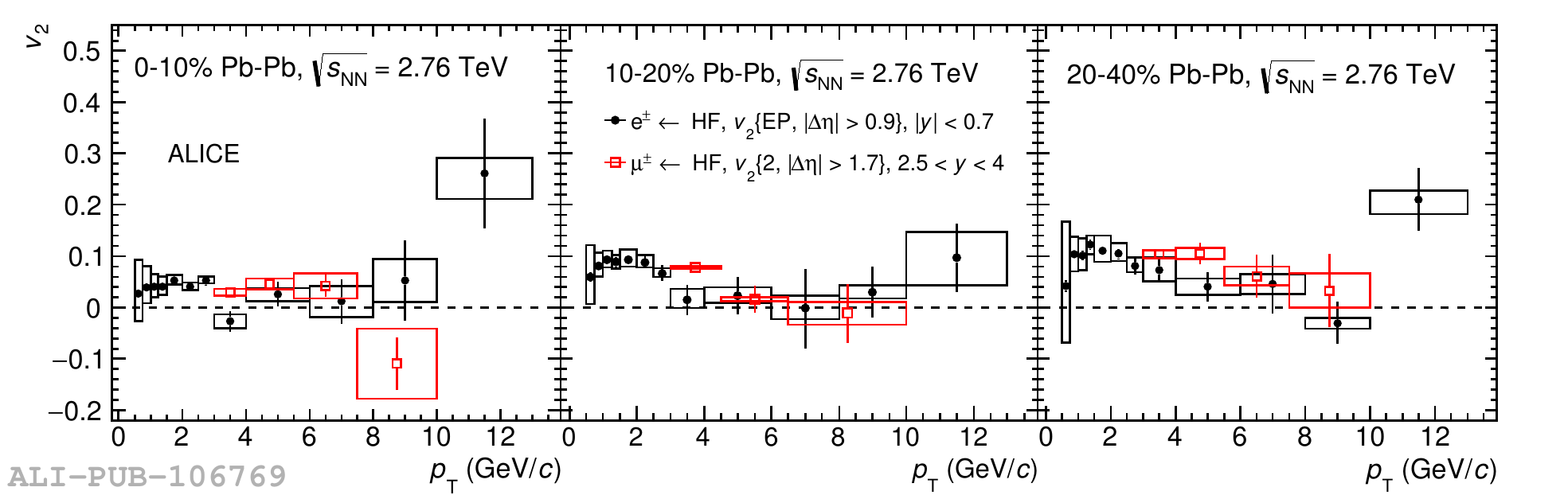}
\caption{Elliptic flow of heavy-flavour decay electrons at mid-rapidity~\cite{HFEv2} ($|y|$ $<$ 0.7) (closed symbols) as a function of $p_{\rm T}$ compared to the elliptic flow of heavy-flavour decay muons at forward rapidity ~\cite{Muon} (2.5 $<$ $y$ $<$ 4) (open symbols) in the 0--10\% (left panel), 10--20\% (middle panel) and 20--40\% (right panel) centrality classes in Pb--Pb collisions at $\sqrt{s_\mathrm{NN}}$~ = 2.76 TeV.}
\label{fig:figure5}
\end{figure*}

The ALICE collaboration also measured the elliptic flow $v_{2}$ of open heavy-flavour hadrons via their hadronic and semi-leptonic decays in Pb--Pb collisions at $\sqrt{s_\mathrm{NN}}$ = 2.76 TeV~\cite{DMESONV2, Muon, HFEv2, DMESONV2Long}.
The measured averaged $v_{2}$  of prompt ${\rm D}^{0}$, ${\rm D}^{+}$ and ${\rm D}^{*+}$  indicates a positive $v_{2}$ in semi-central (30-50\%) Pb--Pb collisions with a significance of 5.7 $\sigma$ for 2 $<$ $p_\mathrm{T}$ $<$ 6 GeV/{\it c} \cite{DMESONV2}.
The anisotropy of prompt  ${\rm D}^{0}$ mesons was measured in the three centrality classes 0-10\%, 10-30\% and 30-50\%, as reported in \cite{DMESONV2Long}. The results show a hint of increasing $v_{2}$ from central to semi-peripheral collisions and are comparable in magnitude to that of inclusive charged particles~\cite{DMESONV2Long, V2Ch}. 
The elliptic flow of heavy-flavour hadron decay electrons at mid-rapidity~\cite{HFEv2} and muons at forward rapidity~\cite{Muon} was measured in the three centrality classes 0-10\%, 10-20\% and 20-40\%. The $v_{\rm 2}$ values are comparable in magnitude as shown in Fig.~\ref{fig:figure5}.
A positive $v_{\rm 2}$ is observed for the heavy-flavour hadron decay electrons in all centrality classes, with a maximum significance of 5.9$\sigma$ in the $p_{\rm T}$ interval 2--2.5 GeV/$c$ in semi-central collisions (20--40\%). At higher $p_{\rm T}$, the measured $v_{\rm 2}$ of heavy-flavour decay electrons exhibits a slight decrease as $p_{\rm T}$ increases, becoming consistent with zero within large uncertainties for $p_{\rm T}$ $>$ 4 GeV/$c$.
A decreasing trend of $v_{2}$ towards central collisions is observed. This is consistent with a final-state anisotropy in momentum space driven by the initial geometrical anisotropy of the nucleons participating in the collision, which increases towards peripheral collisions. 
 
These results indicate that the interactions between the medium and heavy quarks, mainly charm, gives rise to a collective effect measurable via their azimuthal anisotropy, possibly suggesting that charm quarks participate in the collective expansion of the system.

\section{Conclusions}

The results obtained with ALICE using the data from the LHC Run-1 (2010-2013) indicate a strong suppression of heavy-flavour production in central Pb--Pb collisions for $p_\mathrm{T}$ $>$ 3 GeV/c, observed for heavy-flavour decay electrons and muons, for electrons from beauty-hadron decays and for prompt D mesons. From the comparison with p--Pb measurements it is possible to conclude that the suppression observed in Pb--Pb collisions is mainly due to final state effects, i.e. the interaction of heavy quarks with the hot and dense medium, with an indication of a  difference in the D and B-meson suppression at high $p_\mathrm{T}$ as expected by the quark-mass dependence of the energy loss.
The azimuthal anisotropy of D mesons and heavy-flavour decay electrons at mid-rapidity, as well as heavy-flavour decay muons at forward rapidity, measured with ALICE in central and semi-central Pb--Pb collisions at $\sqrt{s_\mathrm{NN}}$ = 2.76 TeV was presented. The results indicate a positive $v_{2}$ in semi-central Pb--Pb collisions and a hint that $v_{2}$ increases  from central to semi-central collisions. The magnitude of the D-meson elliptic flow is similar to the one of charged particles. This suggests a collective motion of low-$p_\mathrm{T}$ heavy-quarks, mainly charm. The $v_{2}$  and $R_\mathrm{AA}$ measurements together can provide constraints to the models about the different energy loss and hadronisation mechanisms.


\nocite{*}
\bibliographystyle{elsarticle-num}
\bibliography{jos}

\begin{thebibliography}{00}

  \bibitem{ALICE2} 
K. Aamodt  {\it et al. }[ALICE Collaboration], Journal of Instrumentation 3 no. 08, (2008) S08002
\bibitem{Radiativea}  
M. Gyulassy and M. Plumer, Phys. Lett. B 243 (1990), 432 - 438
\bibitem{Colla}  
M.Thoma and M. Gyulassy, Nuclear Physics B 351 (1991), 491 - 506
\bibitem{Coale}  
V. Greco, C. Ko, and R. Rapp, Phys. Lett. B 595 (2004), 202 - 208
\bibitem{highptv2}  
E. V. Shuryak, Phys. Rev. C 66 (2002) 027902.
\bibitem{MuonRAA} 
B. Abelev {\it et al. } [ALICE Collaboration], Phys. Rev. Lett. 109 (2012) 112301
\bibitem{Shingopapaper} 
J. Adam {\it et al. } [ALICE Collaboration],  arXiv:1609.07104 [nucl-ex]
\bibitem{DMesonRPBPB} 
J. Adam {\it et al. } [ALICE Collaboration], JHEP 03 (2016) 081
 \bibitem{HFERpPb} 
 J. Adam {\it et al. } [ALICE Collaboration], Phys. Lett. B 754 (2016) 81-93 
\bibitem{DeadCone} 
Y. L. Dokshitzer and D. E.  Kharzeev, Phys. Lett. B 519 (2001), 199 - 206
 \bibitem{DmesonRAACent} 
J. Adam~{\it et al. } [ALICE Collaboration], JHEP 11 (2015) 205
 \bibitem{RAACH} 
B. Abelev {\it et al. } [ALICE Collaboration], Phys. Lett. B 720 (2013) 52 - 62
 \bibitem{RAAPion} 
B. Abelev {\it et al. } [ALICE Collaboration], Phys. Lett. B 736 (2014) 196 - 207
\bibitem{Djordjevic} 
M. Djordjevic, Phys. Rev. Lett. B 737 (2014) 286 - 298 
\bibitem{RAACMS} 
V. Khachatryan~{\it et al. } [CMS Collaboration],  arXiv:1610.00613 [nucl-ex] 
\bibitem{Btoe} 
 J. Adam~{\it et al. } [ALICE Collaboration], arXiv:1609.03898 [nucl-ex]
\bibitem{Btoepp} 
B. Abelev {\it et al. } [ALICE Collaboration], Phys. Rev. D 91 (2015) 012001
 
 
 
\bibitem{DMESONV2} 
 B. Abelev {\it et al. } [ALICE Collaboration], Phys. Rev. Lett. 111 (2013), 102301   
\bibitem{Muon} 
J. Adam~{\it et al. } [ALICE Collaboration], Phys. Lett. B 753 (2016) 41 - 56
\bibitem{HFEv2} 
J. Adam~{\it et al. } [ALICE Collaboration], JHEP 09 (2016) 028 
\bibitem{DMESONV2Long} 
 B. Abelev {\it et al. } [ALICE Collaboration], Phys. Rev. C 90 (2014) 034904   
\bibitem{V2Ch} 
B. Abelev {\it et al. } [ALICE Collaboration], Phys. Lett. B 719 (2013) 18






\end{thebibliography}



\end{document}